\documentclass[12pt,onecolumn,aps,preprintnumbers,prl,showpacs,amsfonts,amsmath,amssymb,superscriptaddress,floatfix]{revtex4}
\pdfoutput=1
\usepackage[]{verbatim}
\usepackage[pdftex]{graphicx}
\usepackage{wrapfig}
\usepackage{subfig}
\usepackage{setspace}
\usepackage{sistyle}
\usepackage{sidecap}
\begin{document}
\title{Increasing the photon collection rate from a single NV center with a silver mirror}
\date{\today}
\author{Niels M\o{}ller Israelsen}\email{nikr@fysik.dtu.dk}
\affiliation{Department of Physics, Technical University of Denmark, Building 309, 2800 Lyngby, Denmark}
\affiliation{These authors contributed equally to this work}
\author{Shailesh Kumar}
\affiliation{Department of Physics, Technical University of Denmark, Building 309, 2800 Lyngby, Denmark}
\affiliation{These authors contributed equally to this work}
\author{Mahmoud Tawfieq}
\affiliation{Department of Physics, Technical University of Denmark, Building 309, 2800 Lyngby, Denmark}
\author{Jonas Schou Neergaard-Nielsen}
\affiliation{Department of Physics, Technical University of Denmark, Building 309, 2800 Lyngby, Denmark}
\author{Alexander Huck}
\affiliation{Department of Physics, Technical University of Denmark, Building 309, 2800 Lyngby, Denmark}
\author{Ulrik Lund Andersen}
\affiliation{Department of Physics, Technical University of Denmark, Building 309, 2800 Lyngby, Denmark}

\begin{abstract}
In the pursuit of realizing quantum optical networks, a large variety of different approaches have been studied to achieve a single photon source on-demand. The common goal for these approaches is to harvest all the emission
from a quantum emitter into a single spatial optical mode while maintaining a high signal-to-noise ratio. In this work,
we use a single nitrogen vacancy center in diamond as a quantum emitter operating at ambient conditions and we
demonstrate an increased photon count rate up to a factor of 1.76 by placing a silver mirror fabricated on the end
facet of an optical fiber near the emitter.
\end{abstract}

\maketitle

\section{Introduction}
In recent years, the negatively charged nitrogen vacancy (NV) center in diamond has become one of the most attractive
solid state quantum emitters due its robustness, stable emission and optically detected magnetic resonance (ODMR).
By exploiting the quality of the NV center, several fundamental phenomena like spin-spin entanglement and detection of
quantum coherent evolution of spin states have been demonstrated ~\cite{Dutt2007,Dreau2013,Dolde2013,Pfaff2014}. This enables the construction of a basic quantum spin register which is an important step towards the construction of a large scale quantum network.
Furthermore, due to its atomic confinement and its long spin coherence time under ambient conditions the single
NV center constitutes an excellent magnetic field sensor both in terms of sensitivity and spatial resolution. Besides the
spin coherence time, the sensitivity of an NV center as a magnetic field sensor is ultimately limited by the amount of
detected photons~\cite{Acosta2013,Jensen2013,Maletinsky2012} and thus a high optical collection rate is of high
importance.

Different methods have been proposed to increase the photon collection rate from an NV center. These approaches
include, among others, the placement of the NV center in a nano-cavity directly processed into the diamond host
material~\cite{Faraon2012}, micro-pillar structures~\cite{Birgit2010,Choy2011}, and solid immersion lenses
(SILs)~\cite{Hadden2010}. In a standard bulk diamond host material with an index of refraction of $n_D = 2.42,$ most of
the light emitted by an NV center remains inside the diamond due to total internal reflection. This limitation could be
overcome by placing detectors on the side of the diamond sample~\cite{Sage2012}, where a 100-times increased signal
could be obtained compared to the standard approach with a confocal microscope. Another approach to increase the photon
collection rate is to place the emitter at or close to the field maximum of a waveguide mode, such that emission
predominantly occurs into the guided mode. This has been shown experimentally for both dielectric~\cite{Schroeder2012,Schroeder2011} and
plasmonic~\cite{Akimov2007, Kolesov2009, Huck2011} waveguides.

Another approach to increase the photon collection rate is to place a metallic mirror in close vicinity of
the emitter \cite{Sandoghdar2011}. This modifies the angular emission pattern of the emitter and  changes the spontaneous decay rate, as
originally demonstrated for Eu$^+$ ions~\cite{Drexhage1970,Amos1997} and very recently for single
NV centers~\cite{Koenderink2013}.

In this article we report on an increase of the photon count rate from three different single NV centers by placing a
silver mirror fabricated on the end-facet of an optical fiber in the vicinity of the emitter. Although the silver
mirror is fluorescing under intense laser light illumination we show that a high signal-to-noise ratio can be obtained
with this approach.

The article is structured as follows. In the first part, we introduce our experimental methods concerning our measurement setup, sample and mirror preparation and measurement approach and we present our measurements demonstrating increased photon collection rates. In the second part, we discuss the two contributions to an increased photon collection rate; the change of the local density of states in the vicinity of the mirror, and the increase of the total collection rate due to reflection of the NV center fluorescence from the mirror and into the mode of collection. We introduce a model for each contribution and compare the net increase of photon collection predicted by these with the experimentally measured value. Finally, we summarize our findings.

\section{Experimental methods}

\begin{figure}[htb]
\includegraphics[width=0.48\textwidth]{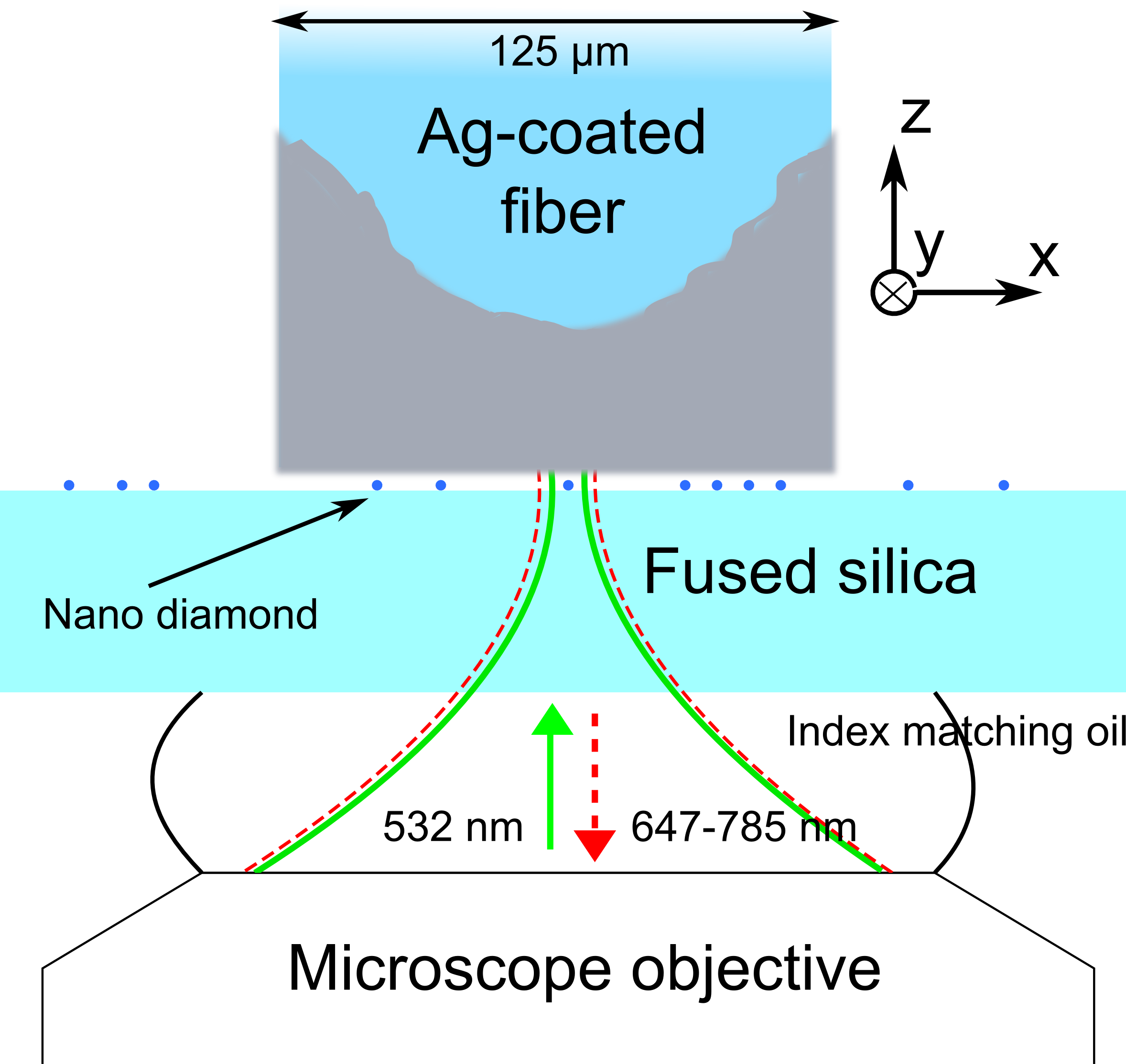} 
\caption{The sample setup. An immersion oil objective focuses the pump beam (green lines) from beneath the substrate onto a selected fluorescing nano-diamond containing an NV center. A part of the fluorescence from the NV center is then coupled back into the objective (red dashed lines). From above a silver-coated fiber can be introduced to increase the coupling into the objective.\label{SMsetup}}
\end{figure}

Our sample was prepared on a plasma cleaned fused silica substrate with a thickness of \SI{0.17}{mm} optimized for a
standard oil immersion microscope objective. After a plasma cleaning process, we spin-coated nano-diamonds with
diameters $< 50 \text{nm}$ (MSY 0-0.05, Microdiamant AG) on the substrate. The density of the nano-diamonds was chosen
to be sufficiently low such that optically active nano-diamonds containing single NV centers could be addressed
individually and characterized by our home built scanning fluorescence confocal microscope. The silver mirror was
prepared on the end-facet of a cleaved optical fiber having a diameter of \SI{125}{\mu m}. The cleaved end facet was
silver-coated with a thickness $>$~\SI{200}{nm} by electron beam evaporation of silver (Alcatel SCM 600).\\
The sample-fiber setup is depicted in Fig.~\ref{SMsetup}. NV centers were excited with a linearly polarized continuous wave laser of wavelength $\lambda_0 = \SI{532}{nm}$ through an immersion oil objective with a numerical aperture NA$ = \text{1.4.}$ After the collection through the same objective, the fluorescence light was band-pass filtered with high transmission in the range between $647 - \SI{785}{nm}$ and
detected by two avalanche photo diodes (APDs) in a Hanbury-Brown and Twiss configuration~\cite{HBT1956}. Our APDs have
dark count rates of 100 counts/s and 500 counts/s, respectively, and, if required, we analyzed the temporal
correlations between the two APDs using a time-to-amplitude converter.\\
After the sample was prepared, the silver coated fiber mirror was introduced by mounting the fiber on a xyz-piezo stage above the substrate as indicated in Fig.~\ref{SMsetup}. To controllably approach with the mirror we imaged the substrate plane with the pump laser by introducing a camera in one of the APD channels. Then, the fiber could be recognized as a second reflecting plane when lowered towards the substrate. Finally, the fiber was tilted in order to cancel out any visible intensity gradient in the reflecting plane to align the fiber end-facet parallel to the substrate plane.\\
To clarify the data acquisition, we shortly explain the measurement approach. The pump was focused on a chosen NV center. The fiber was then scanned in the z-direction in steps of \SI{20}{nm} for a given range. Next, the pump beam was moved in the xy-direction to an area without any emitters. An equivalent second scan was finally performed yielding a background measurement. 

\begin{figure}[htb]
    \centering
    \subfloat[\label{hist}] {\includegraphics[width=0.49\textwidth]{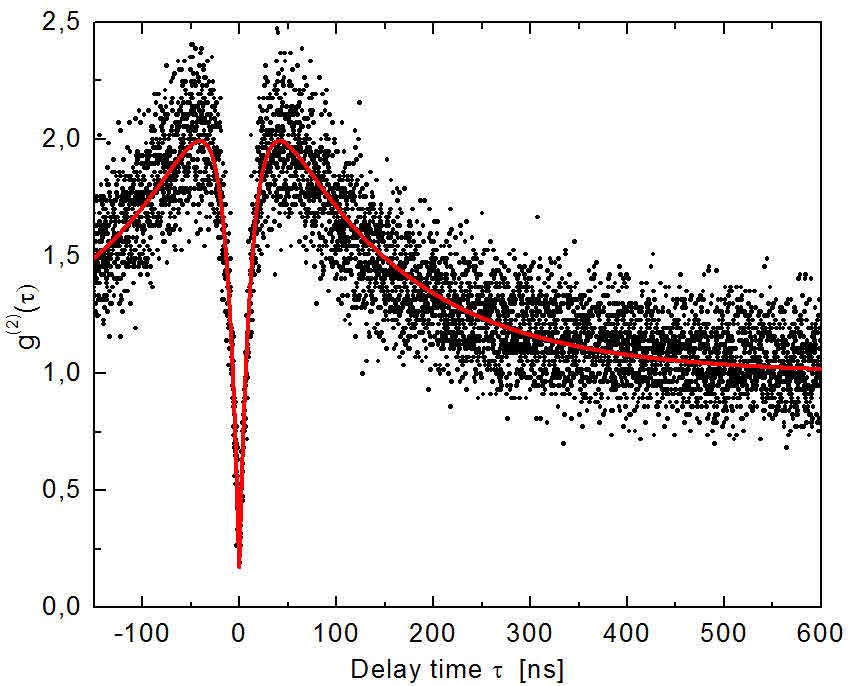}}
    \hfill
    \subfloat[\label{PC}] {\includegraphics[width=0.47\textwidth]{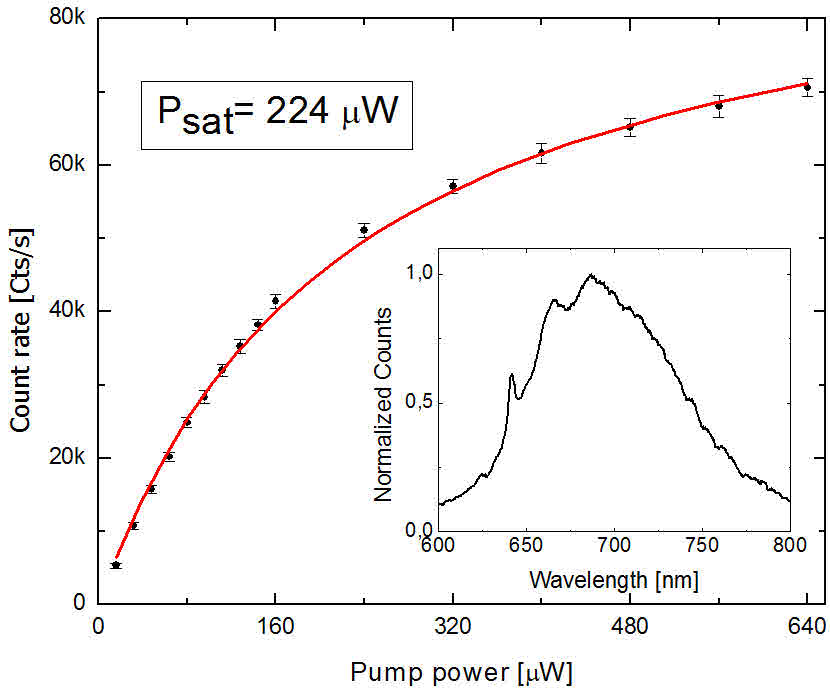}}
    \caption[]{(a) Measured second order correlation function of NVb, obtained for a pump power of \SI{480}{\mu W}. (b) Count rate as function of pump power measured for NVb without a mirror. A fit with a saturation function (red straight line)
    yields a saturation pump power $P_{sat} = 224 \mu W$. Inset: spectrum of NVb.}
\end{figure}

\begin{figure}[h]
\subfloat[\label{BGabs}]
{
\includegraphics[width=0.48\textwidth]{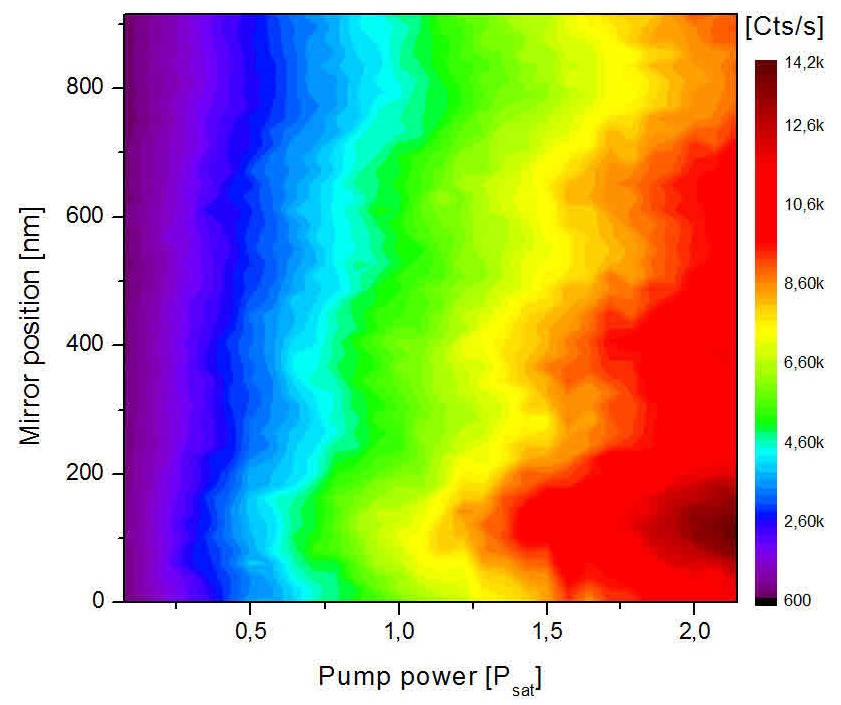}
}\hfill
\subfloat[\label{BGrel}]{
\includegraphics[width=0.48\textwidth]{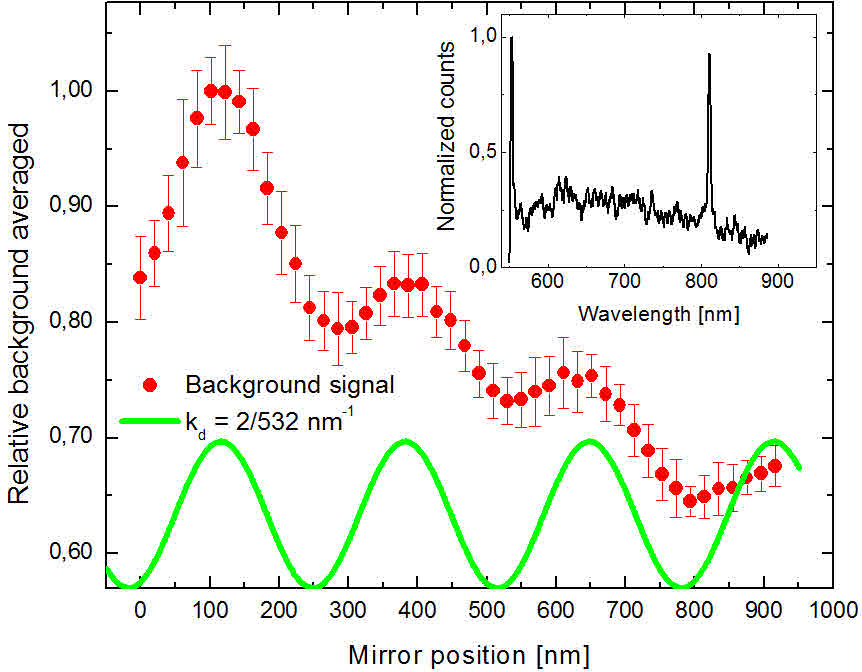}
} \caption[]{(a) Fluorescence map of an electron beam evaporated silver mirror as a function of mirror position and pump power relative to the saturation power $P_{sat}=\SI{224}{\mu W}$ of NVb. (b) Silver mirror fluorescence normalized and averaged with respect to the pump power.
Oscillations with a period corresponding to $532/2$ nm are observed. The green curve is shown as a
guide to the eye and has a period of \SI{266}{nm}. Inset: Spectrum of an electron-beam evaporated silver mirror
obtained at a pump power of \SI{480}{\mu W} and located roughly \SI{500}{nm} from the substrate. Two peaks at \SI{554}{nm} and \SI{811}{nm} are outside the detection band of our setup.\label{noise1}}
\end{figure}

\subsection{Characterization of NV centers} Before introducing the silver mirror we selected and characterized three nano-diamonds
containing single NV centers. For simplicity, these NV centers are referred to as NVa, NVb, and NVc throughout the manuscript. The emission of single photons from the NV centers was verified by measuring the
second-order correlation function $g^{2}(\tau)$. In Fig.~\ref{hist} we plot as an example the measurement result of
$g^{(2)}(\tau)$ for NVb together with a fit based on the model of a three level system yielding $g^{(2)}(0)=0.16$. With the same approach we find $g^{(2)}(0)=0.24$ and $g^{(2)}(0)=0.37$ for NVa and NVc, respectively, proving that all NV centers are single photon emitters as $g^{(2)}(0)<0.5$. The pump power dependent count rate of
NVb is shown in Fig.~\ref{PC}, from which we estimate a saturation pump power of $P_{sat} = \SI{224}{\mu W}$. In a similar fashion we find saturation pump powers of \SI{119}{\mu W} and \SI{150}{\mu W} for NVa and NVc, respectively. The inset in Fig.~\ref{PC} depicts the spectrum of NVb which shows the zero phonon line around \SI{637}{nm} and the phonon side band peak of about \SI{700}{nm}, which is characteristic for the negatively charged NV center at room temperature. Finally, all measurements reported
on in the following were carried out for polarizations of the pump laser yielding the maximum count rate.

\subsection{Characterization of the silver mirror} 
Evaporated silver is known to fluoresce under intense excitation with laser
light having a wavelength of $\lambda_0 = \SI{532}{nm}$ \cite{Peyser2001,Pfaff2013}. Hence, the silver mirror will serve as a source of background noise when
brought within a few \SI{}{\mu m} from an NV center. As mentioned previously, we obtained a background measurement for each signal measure. The recorded count rate resulting from silver fluorescence is presented in Fig.~\ref{BGabs} as a function of mirror position (to be discussed) and pump power. Assuming that the silver fluorescence is spatially uniform and constant
in time, we obtain the background signal important for characterizations of the NV centers, as reported later in the manuscript.
From this measurement it is obvious that the largest mirror fluorescence with a count rate of \SI{14}{kCts/s} is
obtained for small substrate-mirror separations and at high pump power of~\SI{\approx 480}{\mu W}, see Fig~\ref{noise1}a. Due to Fresnel
reflections on the glass-air interface, a standing wave of the pump laser builds up between the silver mirror and the
glass substrate causing periodic changes of the background signal. This is revealed by the plot presented in
Fig.~\ref{BGrel}, where for each pump power the signal was normalized to its maximum value and averaged over all
powers. As expected, the oscillation period corresponds to half the wavelength of the pump laser. In the inset of Fig.~\ref{BGrel}, we plot the spectrum of the silver fluorescence and we clearly see that the silver fluorescence behaves spectrally as a very broadband and uniform noise source within our detection band.

\begin{figure}[htb]

\centering \subfloat[]{ \includegraphics[width=0.48\textwidth]{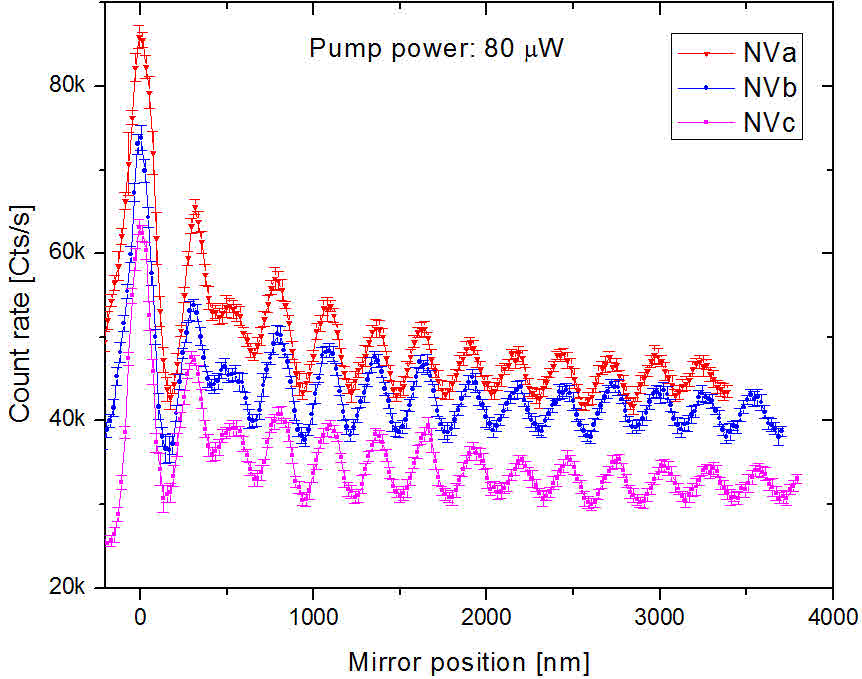}} \hfill
\hfill \subfloat[]{\includegraphics[width=0.48\textwidth]{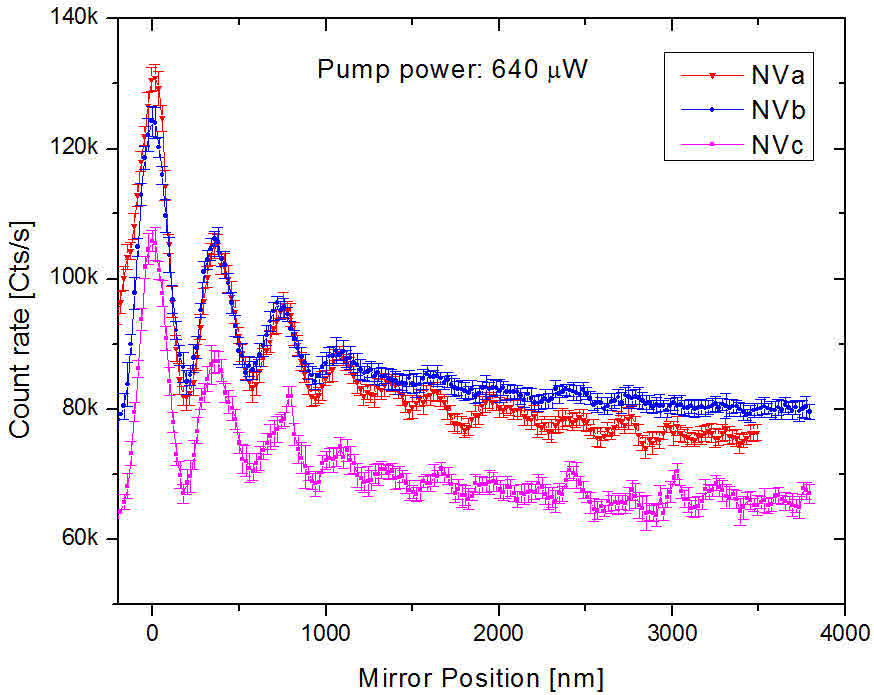} } \caption[]{Fluorescence vs. mirror position
for three NV centers and for two different pump powers. The errorbars mark the statistical uncertainty for ten identical
measurements. All traces are corrected for background fluorescence originating from the silver mirror. The mirror position for the global count rate peaks is labeled \SI{0}{nm}, being approximately $\SI{150}{nm}\pm\SI{54}{nm}$ from mirror-substrate contact.\label{NV345}}
\end{figure}

%%%%%%%%%%%%%%%%%%%%%%% Corrected from here - Niels, 29-05-2014
\subsection{NV centers in the vicinity of a silver mirror} 
Next, we focus on the count rate from the NV centers in the vicinity of a silver mirror. For each NV center the count rate is recorded as a function of mirror-substrate separation~$D$ for
different pump powers. The background corrected count rate obtained from all three NV centers as function of mirror
position is presented in Fig.~\ref{NV345} for a pump power of \SI{80}{\mu W} (a) and \SI{640}{\mu W} (b).
The maximum oscillation peaks are labeled so they occur for a mirror position of \SI{0}{nm}. We have chosen to crop the data for mirror positions smaller than \SI{-200}{nm} since the mirror at this point is touching the substrate which is recognized by a sudden an-harmonic behavior of the fluorescence with decreasing mirror position in each mirror scan measurement. 

For the pump power $P=\SI{80}{\mu W}$ and substrate-mirror separations in the range 1000-\SI{4000}{nm}, significant count rate oscillations are visible, with periods corresponding to (or close to) half the pump laser wavelength $\frac{532}{2}$ nm. This occurs due to the standing field created from interference between the pump beam approaching the mirror and the reflected pump beam traveling in the opposite direction. In contrast to this, when $P
> P_{sat}$ (cf. Fig.~\ref{NV345}b) oscillations of the count rate corresponding to $\frac{532}{2}$ nm are barely visible
since in this range the NV center level populations are almost independent on small changes of the pump power.

All measurements presented in Fig.~\ref{NV345} show that for substrate-mirror separations below \SI{1000}{nm}, strong
count rate oscillations occur. For $P > P_{sat}$, the period of these oscillations dominate the pump associated oscillations and appear to be approximately \SI{350}{nm} corresponding to the
maximum of the phonon broadened sideband peak in the NV centers emission spectrum. Taking the maximum count rate in the
high excitation power limit relative to the count rate without a mirror, enhancement by a factor of $1.44\pm 0.040$, $1.76\pm 0.045$, and $1.57\pm 0.036$ are observed for NVa, NVb, and NVc, respectively.

\begin{figure}[htb]
\subfloat[\label{SigAbs}]
{
\includegraphics[width=0.47\textwidth]{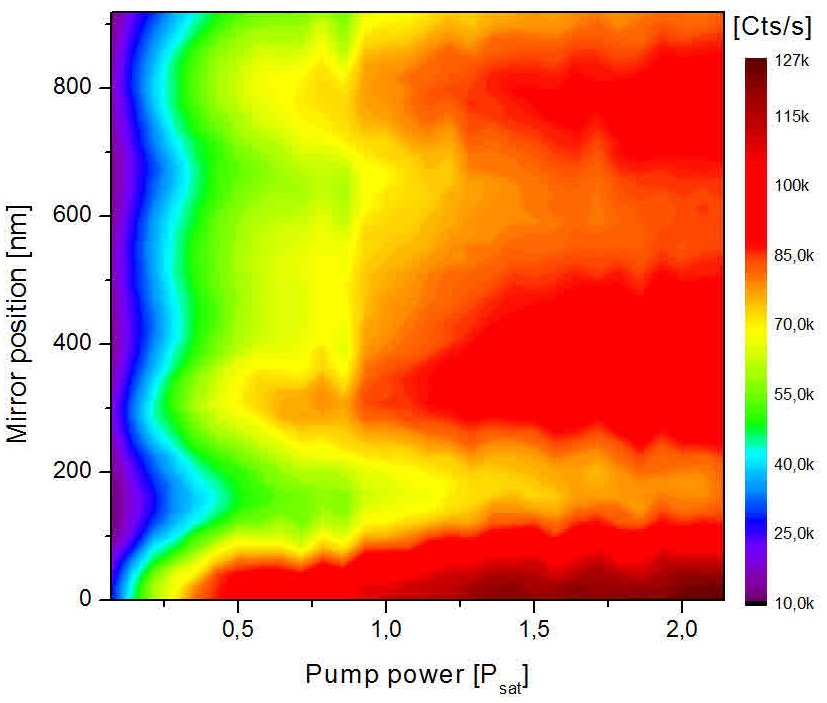}
}\hfill
\subfloat[\label{SigRel}]{
\includegraphics[width=0.49\textwidth]{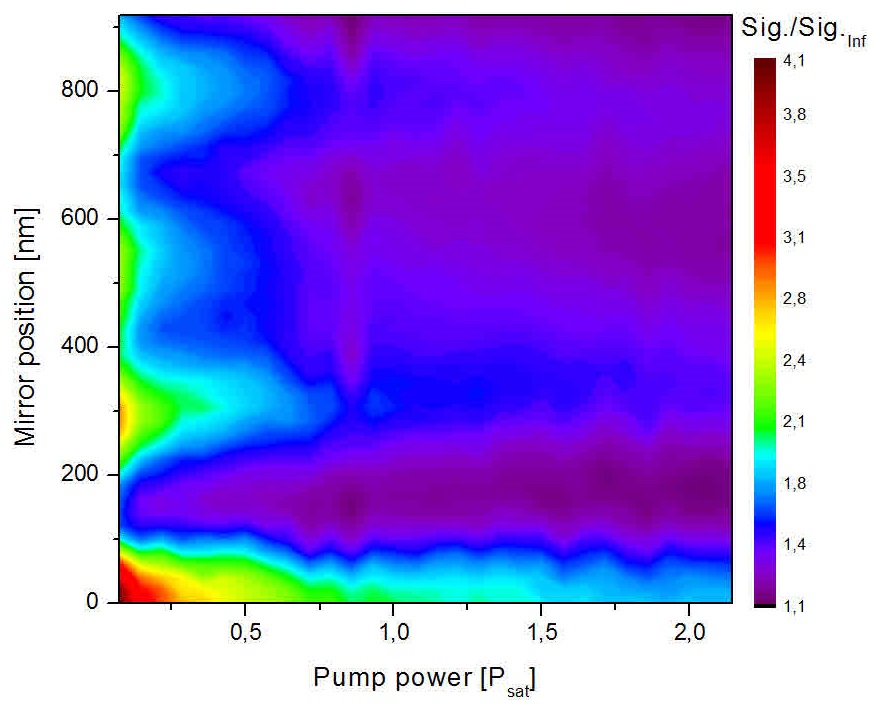}
} \caption[]{(a) Background corrected count rate of NVb as a function of mirror position and pump power. (b) The same
as in in (a), but normalized to the count rate without the mirror $(D=\infty)$ presented in Fig.~\ref{PC}.\label{color1}}
\end{figure}

To investigate the underlying dependencies more thoroughly, we scan for NVb the excitation power in steps of
\SI{100}{\mu W} and record the count rate as a function of substrate-mirror separation. The result of this background corrected measurement
is presented as a color map in Fig.~\ref{SigAbs} and the background corrected result normalized to the counts without
mirror (cf. Fig.~\ref{PC}) is shown in Fig.~\ref{SigRel}. The largest enhancement is obtained for small NV center-mirror separations
yielding a factor of around $4$ at a low  pump power and a factor of around $1.75$ for a high pump power. In accordance
 with the measurements shown in Fig.~\ref{NV345}, for $P \gtrsim P_{sat}$ the oscillations corresponding to a $\frac{532}{2}$ nm period
decrease and the signal oscillates with longer periods corresponding to the maximum of the NV center spectrum.
\\

\begin{figure}[h!]
\includegraphics[width=0.6\textwidth]{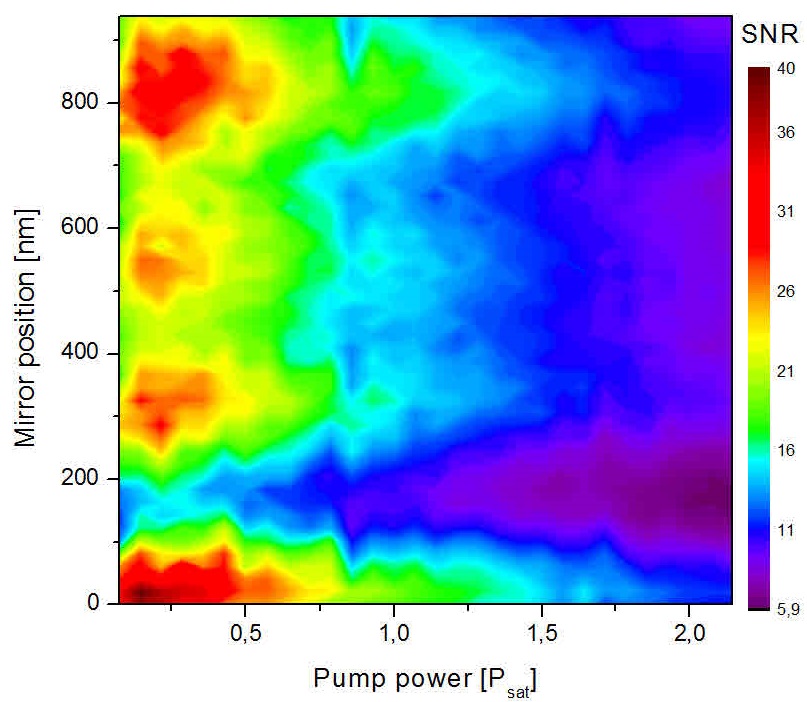}
 \caption[]{The signal-to-noise ratio determined from the ratio between background corrected absolute counts (cf. Fig.~\ref{SigAbs}) and mirror fluorescence (cf. Fig.~\ref{BGabs}).\label{StNabs}}
\end{figure}

When measuring specific parameters, one can not rely on a subsequent subtraction of background fluorescence and hence the
signal-to-noise ratio (SNR) is most relevant. The SNR is obtained as the ratio between the background corrected count
rate (cf. Fig.~\ref{SigAbs} for NVb) and the count rate from the silver mirror (cf. Fig.~\ref{BGabs}). It is plotted in
Fig.~\ref{StNabs} for NVb. At a low pump power the highest SNR is measured with a value of 40, while for high pump
powers the SNR decreases to around 12 due to a largely increased background fluorescence of the silver mirror.\\

The background fluorescence from the silver mirror limiting the SNR can be a result of many effects such as
roughness of the cleaved fiber facet, the grain type and size for the deposited silver and silver-composites like
silver-oxide and silver-sulfide forming on the silver surface in ambient air \cite{Pett1995,Burge1969,Bennett1970}. A careful optimization of the silver deposition can however remedy the low SNR. We tested an alternative deposition method which was a wet chemical deposition~\cite{Radu1}. Despite that this method provided uniform smooth silver deposition independent on surface orientation, we found that the method introduced previously provided silver layers which fluoresced less and in a more stable manner under \SI{532}{nm} laser illumination.\\
The SNR will of course also intrinsically depend on the brightness of the NV center as
well as the quality and hence the amount of fluorescence stemming from the mirror. The brightness of the NV center
depends both on the geometry of the host diamond as well as the local charge environment. While the first feature
inflicts collection efficiency and the spontaneous emission rate, the second feature inflicts the disruption of the electronic
excitation-decay cycle which can be a result of charge traps in the diamond bandgap due to additional lattice defects
and/or surface states which are significant for nano-diamonds~\cite{Wolters2012, Beha2012}.\\

\begin{figure}[h!]
\centering
\includegraphics[width=0.6\textwidth]{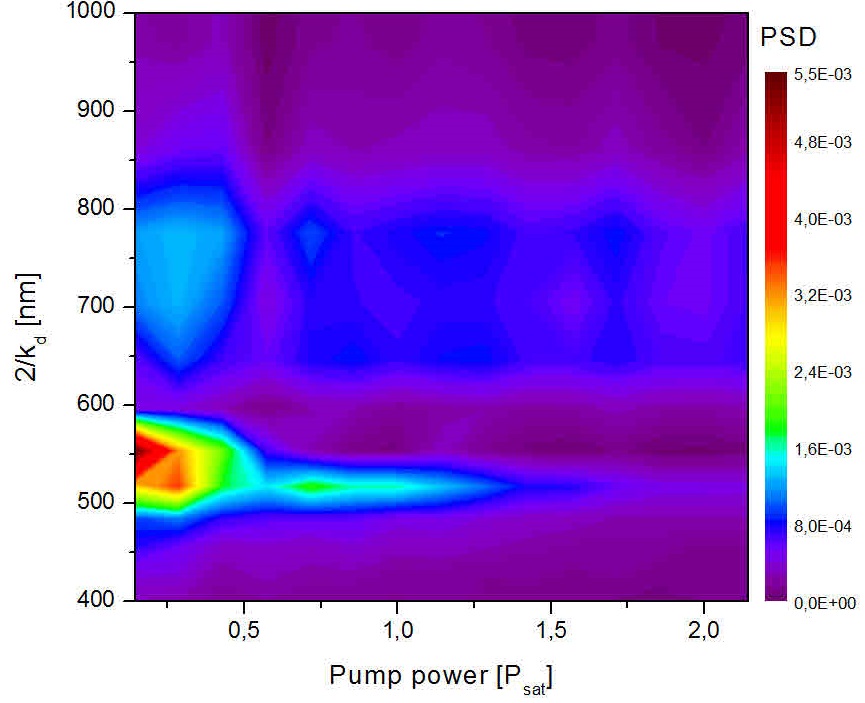}
\caption[]{Power Spectral density (PSD) of spatial components obtained via a fast Fourier transform with respect to the mirror separation. We plot $2/k_{d}$ instead of $k_{d}$ in order to directly illustrate the underlying spatial component.\label{fourier}}
\end{figure}

To resolve the different frequency components contributing to the count rate oscillations at various pump
powers, we performed a fast Fourier transform (FFT) with respect to the substrate-mirror separation of the background
corrected relative signal shown in Fig.~\ref{SigRel}. The power spectral density obtained from the FFT is presented in
Fig.~\ref{fourier}. Instead of plotting the spatial frequency component $k_d$ we re-scaled the axis and plotted $2/k_{d}$
to better illustrate the one-to-one wavelength correspondence. Two significant $2/k_{d}$-components are visible in
Fig.~\ref{fourier}: one at around \SI{532}{nm} confirming the dependence on the pump laser wavelength and a second one
at around \SI{700}{nm}. It is clearly visible that the \SI{532}{nm} dependence decreases for an increasing pump power
and the contribution almost vanishes at a power of $P\gtrsim2P_{sat}$. In contrast to this, the peak centered around
$2/k_{d} = 700 \text{nm}$ is almost independent of the pump power and clearly dominates the spectrum for $P_{exc} >1.5\times
P_{sat}$. The spatial component around $700 \text{nm}$ corresponds to the maximum of the phonon broadened side band peak of the
NV center's emission spectrum (see inset in Fig.~\ref{PC}) and therefore we conclude that these oscillations occur due to an enhancement of the
NV center's spontaneous emission rate in the vicinity of the mirror.

\section{Modeling the collected fluorescence}
\begin{figure}[h!]
\centering
\includegraphics[width=0.6\textwidth]{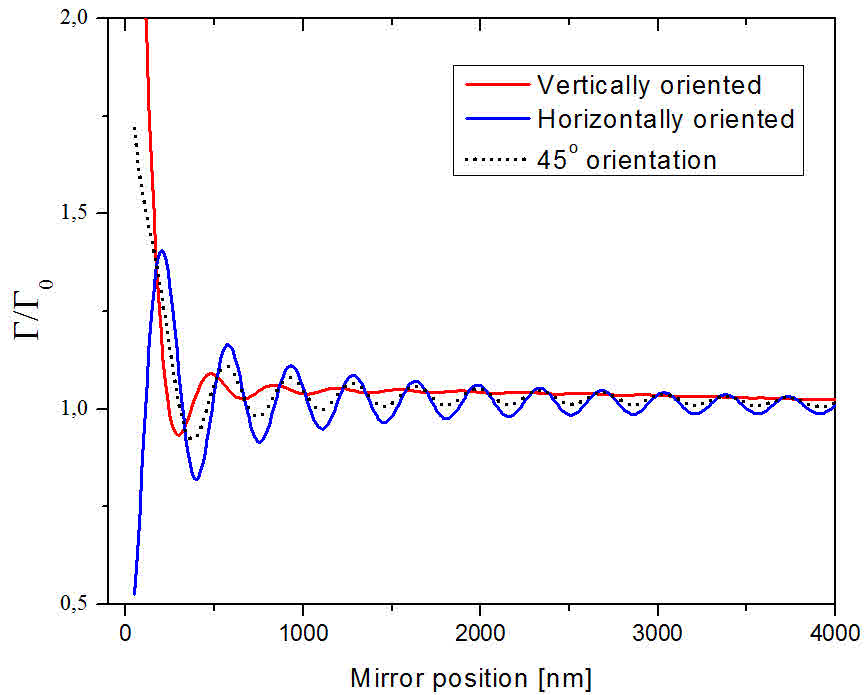}
\caption[]{Relative spontaneous decay rate of a dipole emitter in the vicinity of a silver mirror for different dipole
orientations with respect to the mirror surface. The emission wavelength of the dipole was chosen to be
$\lambda_0=\SI{700}{nm}$ as it corresponds to the peak emission wavelength of an NV center at room temperature. The
quantum efficiency of the emitter was assumed to be one in these calculations. \label{Decayrate}}
\end{figure}
First we consider the photon collection increase due to the Purcell enhancement.
When a mirror is introduced in the vicinity of a dipole emitter the rate of spontaneous emission is changed due to a
variation of the local density of states. The enhancement of an emitter's spontaneous emission decay rate is generally
known as the Purcell effect~\cite{Purcell1946}. In the vicinity of a mirror this effect can be interpreted as a driven
oscillating dipole, where the emitted field interferes with the dipole upon reflection on the mirror and thus changes
the dipole transition~\cite{Drexhage1970,Chance1978}. Our simulation is carried out for a dipole emitting light of $\lambda_0 = \SI{700}{nm}$ representing the NV center phonon sideband and recognized as a present spatial frequency in Fig.~\ref{fourier}. The relative change of the spontaneous decay rate $\Gamma/\Gamma_0,$ where $\Gamma_0$ is the decay rate in a homogeneous environment, versus the distance to a metallic mirror is plotted in Fig.~\ref{Decayrate} for three different dipole orientations with respect to the mirror plane. Here we see a significant difference in the $\Gamma/\Gamma_0$ vs. mirror position behavior depending on the dipole orientation.

The intrinsic property of the NV center having two orthogonal dipole transitions spanning a plane ensures that there will always be a non-zero projection onto the mirror plane. To describe the complete relative weighting of each of these dipole orientations, one needs to take into account both the projection of the pump laser polarization onto each dipole transition as well as the projection of each dipole transition onto the mirror. This is, however, beyond the scope of this paper.

Comparing with Fig.~\ref{NV345}b, where $P>2P_{sat}$ for all three NV centers, we find indications that all of them show similarities with the horizontally oriented emitter recognized by clear oscillations growing in amplitude for an approaching mirror within $\SI{1}{\mu m}$ and a maximum decay rate terminating in a significant drop when approaching $D=0$.  

Assuming that the maximum fluorescence peak in Fig.~\ref{NV345}b indeed corresponds to the maximum decay rate peak in Fig.~\ref{Decayrate} we should experimentally obtain a fluorescence increase of about 1.4 contributed by the Purcell enhancement at an emitter-mirror separation of about \SI{200}{nm}.

\begin{figure}[h!]
\centering
\includegraphics[width=0.98\textwidth]{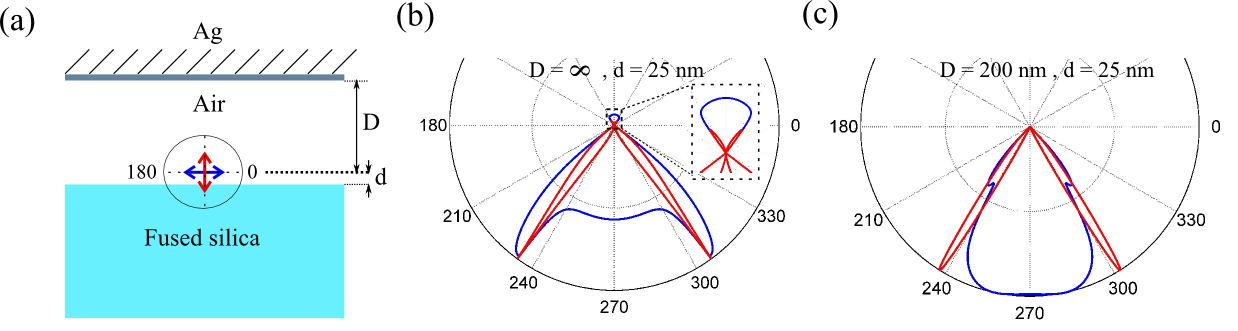}
\caption[]{(a) A model of the experimental configuration: A horizontal (vertical) dipole labeled with a blue (red) arrow is located $d$ above a silica substrate ($n_{SiO_2} = 1.46$). A silver mirror ($n (\lambda_0 = 700nm)  = $0.16761
+ 4.2867i$)$ with parallel alignment is approached to the emitter with a separation $D$. Normalized angular emission
pattern for the configuration shown in (a) with $D=\infty$ (b) and $D=\SI{200}{nm}$ (c). In (b) and (c) $d=\SI{25}{nm}$ and an emission wavelength of $\lambda_0=\SI{700}{nm}$ was used for the calculation.\label{model1}}
\end{figure}

We now consider the increase of photon collection due to reflection of the NV center fluorescence from the mirror and into the mode of collection which we for simplicity call the geometrical factor. When considering a dipole emitter resting near the surface of a substrate, optical near field effects will determine
the angular emission pattern of the dipole \cite{Lukosz1,Lukosz2,Lukosz3}. The angular emission pattern then depends on the
refractive indices of the substrate and the emitter's host material as well as the distance and the orientation of the dipole
with respect to the interface. For multilayer structures the dipole emission pattern can be modeled by a Fourier
integration technique of transmitted and reflected plane waves~\cite{Neyts1998,Chen2007,Chen2011}. Exploiting this technique, we simulate
the emission pattern of a dipole located in air \SI{25}{nm} above a silica substrate with an index of refraction
$n_{SiO_2} = 1.46,$ as depicted in Fig.~\ref{model1}a. We again choose the dipole transition frequency to correspond to a wavelength of $\SI{\lambda_0 = 700}{nm}$ according to the spatial component in Fig.~\ref{fourier} corresponding to the phonon side band peak of the NV center. Vertical (horizontal) orientation refers to perpendicular (parallel) alignment
of the dipole axis to the plane of the substrate. The calculated emission pattern for this case is presented in
Fig.~\ref{model1}b. It is clear that independently of the dipole orientation a fraction of
the emitted light will escape into the upper half space and thus can not be collected by a microscope objective
placed below the glass substrate. With an index-matched high NA microscope objective of NA = 1.4
light within a half angle of $\SI{73.5}{^\circ}$ can be collected which yields a total collection efficiency of 0.79
(0.83) for a horizontally (vertically) aligned dipole. If a silver mirror with $n_{Ag} (\lambda_0=700 nm) = 0.16761
+ 4.2867i$ is positioned \SI{200}{nm} above the emitter, the dipole pattern drastically changes, as illustrated
by the plot in Fig.~\ref{model1}c. The total collection efficiency then increases to 0.99 (0.96) for a horizontally
(vertically) aligned dipole. This corresponds to a geometrical factor in the photon collection being in the range 1.16-1.25 depending on the dipole orientation.\\
Note that the simulation only takes angular dependence into account. This means that light originating from elsewhere in the xy-plane relative to the emitter as a result of multiple scattering between the mirror and the substrate is pictured as stemming from the xy-coordinate of the emitter. But since the lens diameter of the objective $D_{lens}\gg D+d$, this phenomenon will not affect our collection and the emission pattern of Fig.~\ref{model1}c is representative.\\

Considering both the Purcell factor and the geometrical factor, an increase of the total photon count rate up
to a factor of 1.75 is expected for a horizontally aligned dipole separated by approximately $D=\SI{200}{nm}$ from the mirror.

\section{Discussion}

Multiplying the geometrical factor and the Purcell
factor yields a total factor ranging from 1.62 to 1.75 depending on the dipole orientation. The measured collection enhancement of $1.76\pm 0.045$ for NVb fits well this range suggesting that the mirror has reached a distance of approximately \SI{200}{nm} from the emitter.
Since the same mirror, and thus the same mirror alignment, was used for all three NV centers, all of them are probed
for the same mirror distances outlining the oscillation peak nearest to the mirror as depicted in Fig.~\ref{NV345}. The
collection enhancement factors of NVa and NVc are not represented in the $1.62-1.75$ interval. 
Since the models presented only consider single dipole emitters we note that this simplification of the NV center can contribute to the deviation. Furthermore, effects stemming from the unknown individual diamond host geometry are not considered.\\

In case of real-time recordings, where post-subtraction of the background fluorescence is not an option, the ratio between the emitter signal and the mirror background becomes critical. This includes temporal
correlations such as measuring the auto-correlation function $g^{(2)}(\tau)$ or demonstrating the Hong-Ou-Mandel effect. On the other
hand, introducing a mirror for collection enhancement for ODMR measurements would provide a higher noise floor but
would however simultaneously increase the contrast thus improving the distinguishability between e.g. different spin
states due to the increase in collection efficiency \cite{Acosta2013,Jensen2013,Maletinsky2012}.\\
Choosing a different emitter enabling an alternative spectral filtering might increase the SNR if the primary
emitter signal is confined to a narrow band in contrast to the NV center. For this purpose quantum emitters like the silicon-vacancy defects in diamond~\cite{Neu2011,Neu2012}, vacancy defects in silicon
carbide~\cite{Castelletto2013} and nano-structured quantum dots~\cite{Shields2007} can be suited for increasing the collection efficiency using a mirror.\\

Examples of experiments which can benefit from our photon collection enhancement method are experiments with opto-mechanical membranes \cite{Faraon2012,Preeti2014}, and ODMR detection schemes \cite{Acosta2013,Jensen2013,Budker2007}.

\section{Summary}
In summary, we have demonstrated a method for increasing the photon collection rate consisting of a silver coated standard optical fiber. The collection enhancement was demonstrated for three different NV centers where we found count rate increases of $1.44\pm 0.040$, $1.76\pm 0.045$, and $1.57\pm 0.036$. We thoroughly investigated the pump power dependence for one of the NV centers finding a SNR of 12 when saturating the NV center. We introduced two theoretical models which when combined predicted a maximum enhancement of 1.75. We finally discussed our method as a simple way to increase the collection rate for experiments with quantum emitters.

\end{document}